\documentclass[graybox]{svmult}

\usepackage{mathptmx}       
\usepackage{helvet}         
\usepackage{courier}        
\usepackage{type1cm}        
\usepackage{makeidx}         
\usepackage{graphicx}        
\usepackage{multicol}        
\usepackage[bottom]{footmisc}

\makeindex             

\newcommand{\kmsMpc}{km~s$^{-1}$~Mpc$^{-1}$}

\newcommand{\Mpch}{$h^{-1}$~Mpc}
\newcommand{\hMsun}{$h^{-1}\;\mathrm{M}_\odot$}

\begin{document}

\title*{Robust, data-driven inference in non-linear cosmostatistics}
\author{Benjamin D. Wandelt \and Jens Jasche \and Guilhem Lavaux}
\authorrunning{ Wandelt et al.}
\institute{Benjamin D. Wandelt \at 
UPMC Univ Paris 06, UMR7095, Institut d'Astrophysique de Paris, F-75014, Paris, France \at
CNRS, UMR7095, Institut d'Astrophysique de Paris, F-75014, Paris, France\at
 Departments of Physics and Astronomy, University of Illinois at
  Urbana-Champaign, Urbana, IL 61801, USA\at
\email{wandelt@iap.fr}\and 
Jens Jasche \at
Argelander-Institut f\"{u}r Astronomie , Auf dem H\"{u}gel 71,  D-53121 Bonn, Germany
\and
Guilhem Lavaux \at 
Department of Physics and Astronomy, University of Waterloo, 200 University Avenue West, Waterloo, Ontario, Canada, N2L 3G1
\at Perimeter Institute for Theoretical Physics, 31 Caroline Street North, Waterloo, Ontario, N2L 2Y5}

%
%
\maketitle

\abstract*{We discuss two projects in non-linear cosmostatistics applicable to very large surveys of galaxies. The first is a Bayesian reconstruction of galaxy redshifts and their number density distribution from approximate, photometric redshift  data. The second focuses on cosmic voids and uses them to construct \textit{cosmic spheres} that allow reconstructing the expansion history of the Universe using the Alcock-Paczynski test. In both cases we find that non-linearities  \textit{enable} the methods or \textit{enhance} the results: non-linear gravitational evolution creates voids and our photo-z reconstruction works best in the highest density (and hence most non-linear) portions of our simulations.}

\abstract{We discuss two projects in non-linear cosmostatistics applicable to very large surveys of galaxies. The first is a Bayesian reconstruction of galaxy redshifts and their number density distribution from approximate, photometric redshift  data. The second focuses on cosmic voids and uses them to construct \textit{cosmic spheres} that allow reconstructing the expansion history of the Universe using the Alcock-Paczynski test. In both cases we find that non-linearities \textit{enable} the methods or \textit{enhance} the results: non-linear gravitational evolution creates voids and our photo-z reconstruction works best in the highest density (and hence most non-linear) portions of our simulations.}

\section{What is cosmostatistics?}
\label{sec:intro}
Cosmostatistics is the discipline of using the departures from homogeneity observed in astronomical surveys to distinguish between cosmological models. It therefore plays a central role in the cosmological agenda for the coming decade, which is to 
\begin{itemize}
\item learn about the cosmic beginning;
\item understand the cosmic constituents, in particular Dark Matter and Dark Energy; and
\item understand cosmological evolution from initial seed perturbations to current observations
\end{itemize}
One of the challenges for cosmostatistics is that any given observable (maps of the cosmic microwave background, galaxy survey, etc.) is informative about all these goals in some way.

\begin{figure}[]
\begin{center}
\includegraphics[scale=.33]{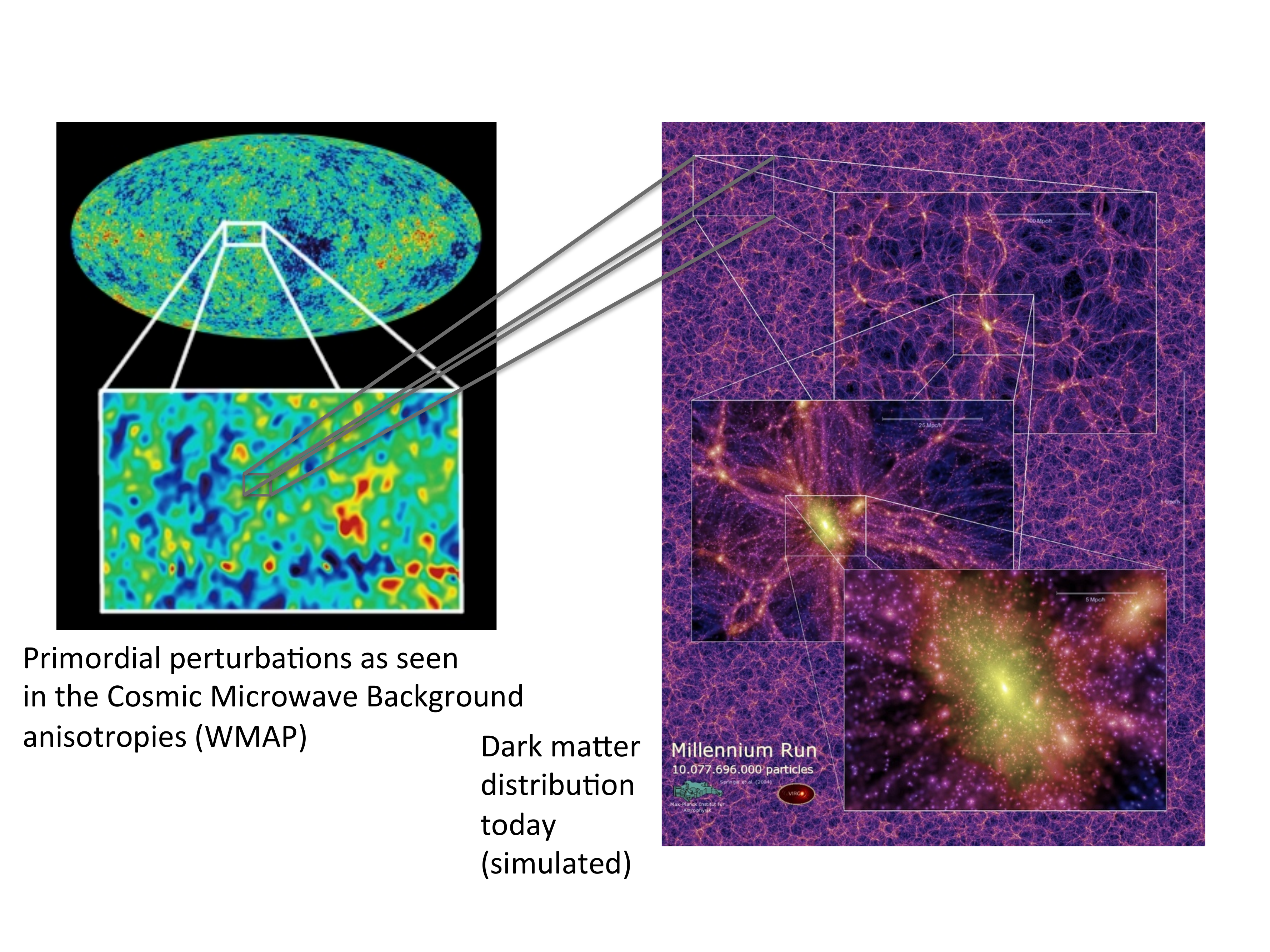}
\end{center}
%
%
\caption{Cosmostatistics uses the stochastic departures from homogeneity on all observable scales to distinguish between cosmological models.}
\label{fig:1}       
\end{figure}

We are fortunate to live in a time when the cosmic microwave background (CMB)is being mapped with high precision from space (by the WMAP \cite{WMAP7_K11} and Planck \cite{planck} missions), and ground-based   and space-based missions are mapping out sizable fractions of the observable Universe in exquisite detail and in three dimensions, across large swaths of the electromagnetic spectrum. Between these two approaches we expect the CMB to have much more signal on very large scales, whereas in principle, probes of density \textit{should} win overall, simply since there are vastly more modes in a three-dimensional data set which greatly reduces sample variance.

How do we realize the immense promise of large scale structure surveys for constraining cosmological models? A number  of known and unknown systematics stand between where we are now and the dream of accessing the vast number of perturbation modes sampled by tracers of the underlying density field. Many of these systematics complicate the relationship between the distribution of tracers and the mass distribution we would actually like to probe. 

These complications arise either due to the intricate physics of galaxy formation or through incomplete information in the data (e.g. having access only to approximate  photometric redshift information instead of the much more expensive spectroscopic redshifts). In addition, the mass density has undergone non-linear dynamical evolution on length scales less than $\sim 20 Mpc/h$, which has coupled the perturbation modes in ways that are non-trivial to model. Non-linear mode coupling erases  information  that the mode amplitudes carried about the state of the early Universe from whence they arose. On the largest scales the limits are set by causality and hence the finite volume of the observable Universe.

Most people would agree on the impracticality of incorporating fully non-linear gravitational evolution into cosmological inference, let alone a fully physical model of galaxy formation. So the challenge is to find ways of looking at the data that are robust to these systematics. 

When it comes to dealing with incomplete information, the challenge is to produce a joint analysis with uncontroversial prior information that allows reconstituting some of the information that has not been captured in the data.

In this talk we will highlight two recent papers  which give examples of these two approaches. In one case \cite{JW11}, we develop a Bayesian approach to improving photometric redshift estimates (and simultaneously estimate the density of the tracers). The prior information we assume to achieve this information recovery is local isotropy of the tracer distribution. 

In the second paper \cite{LW11} we define a new observable to prove the physical properties of dark energy: stacked voids. In this case we choose a very specific pre-processing step to extract features of the data which should be  robust to galaxy bias and to non-linearity. The approach explicitly projects out the details of the tracer distribution in the non-linear density field to obtain nearly spherical objects that nearly co-move with the expansion which serve as the basis of a powerful and purely geometrical test of the expansion history of the Universe. Again, local isotropy underlies this approach which posits that underdense regions are not preferentially oriented with respect to an observer's line of sight.

\section{Bayesian inference from photometric redshift surveys}
\label{sec:BayesianLSS}
\begin{figure}[]
\begin{center}
\includegraphics[scale=.67]{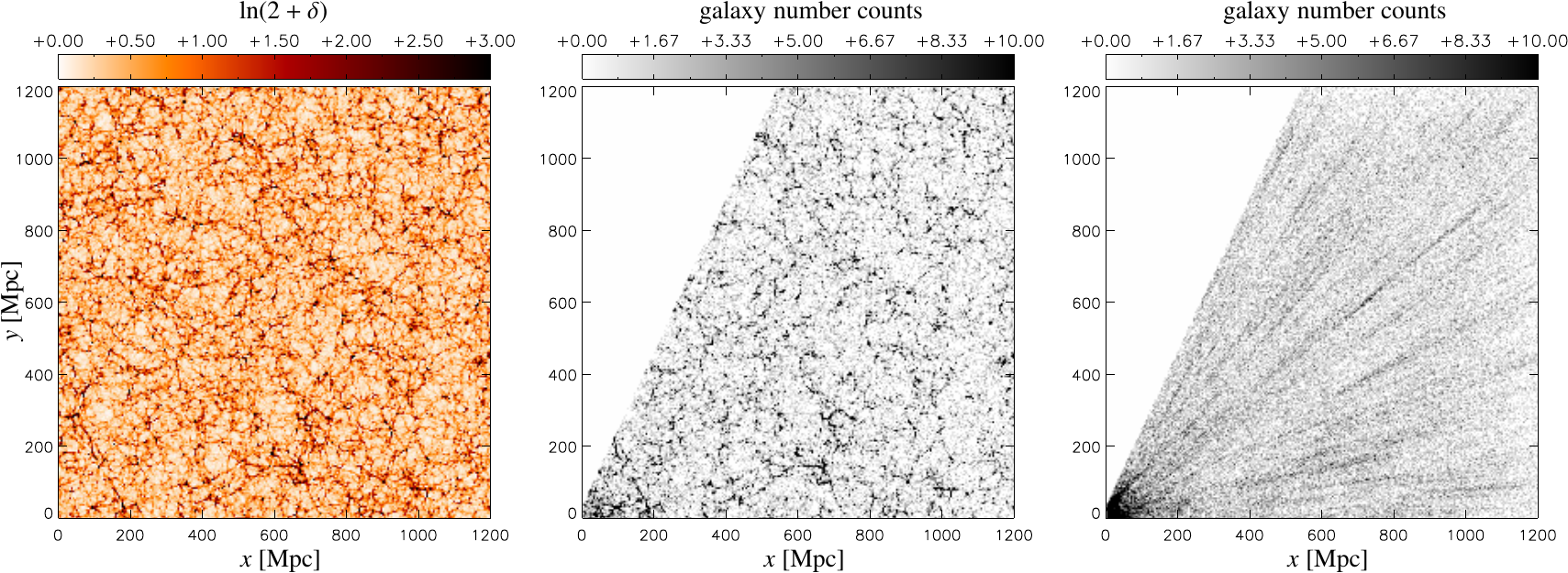}
\end{center}
%
%
\caption{From an n-body simulation to the simulated photo-z survey: the particle density in the simulation (left), after application of the mask (center), and after simulation of photo-z uncertainties (right).}
\label{fig:sim}       
\end{figure}

\index{photometric redshift}\index{Bayesian inference}
The vast majority of ongoing and future surveys (CFHTLS, DES, Pan-STARRS, LSST) are or will be photometric. This is a simple consequence of the cost of taking a galaxy spectrum with current technology. Photometric redshift errors of $\Delta z~0.03$, the current state-of-the-art, translate into smearing along the line of sight on scales of $\sim 200 Mpc$. Such errors are not detrimental to certain kinds of science but will cause any structure smaller than 100 Mpc to be wiped out, as illustrated in Figure~\ref{fig:sim}.

Looking at the trivial density estimate calculated binning  photometric tracers shown in Figure \ref{fig:sim} it is immediately clear that the line-like finger-of-god artifacts introduced by photo-z smearing are very recognizable, since they break local isotropy, a core element of our cosmology. Since they stand out so visibly, we wondered if they could be removed.

In the following we will often refer to the tracers as galaxies, but the nature of the tracer is of no importance to the functioning or implementation of the algorithm. 

\subsection{A simple model of a photo-z catalogue}
First we build a hierarchical model for the distribution of tracers. A simple approach is to consider the points an inhomogeneous  Poisson process. The intensity function of the Poisson process is the underlying number density field, which in turn is a correlated, statistically isotropic, log-normal random field. For the purposes of this exercise we will assume that the correlation function (or equivalently the power spectrum $P(k)$) is known. Relaxing this assumption will be subject of a future study.

The third level in the model hierarchy: photo-z distortions modify the galaxy positions along the radial lines of sight. It is assumed that the redshift uncertainties are specified in terms of a pdf for each tracer. These photo-z pdfs are assumed to be the output of an earlier analysis step which uses any information available, except the spatial distribution of the tracers in the catalog. All photometric information for the galaxy including any morphological features that can be discerned in the images are fair game.

\subsubsection{Implementation}
This hierarchical model can be straightforwardly implemented. The challenge is to explore the posterior density in an efficient manner since the parameter space is enormous: approximately 16 million parameters for the number density and 20 million galaxy redshifts.  We choose a block Gibbs-Metropolis-Hastings sampling approach with the following steps:
\begin{description}
\item[Sample the number density given the current galaxy redshifts.] We draw from the conditional posterior of the number density assuming that the current ``guess'' of the galaxy redshift is correct. This is a solved problem\cite{JASCHE2010HADESMETHOD}; it uses a Hamiltonian sampling approach to update the number density field using the galaxy positions and incorporating the correlated log-normal prior.
\item[Sample the galaxy redshifts given the number density.] The redshift posteriors for the galaxies are conditionally independent given the number density field. This feature allows parallelizing this step over the number of galaxies. Each galaxy performs one step of a Metropolis-Hastings Markov Chain Monte Carlo along the line of sight. The conditional posterior for each galaxy is the product of the input photo-z pdf for this galaxy and the number density. 
\end{description}
Conditional independence is the key feature that allows  this algorithm to scale to tens of millions of galaxies. From the perspective of the message passing paradigm of Bayesian inference, the number density field communicates the information about all the other galaxies to each individual one.

\subsection{Results}
\label{sec:results1}
\begin{figure}[]
\begin{center}
\includegraphics[scale=0.67]{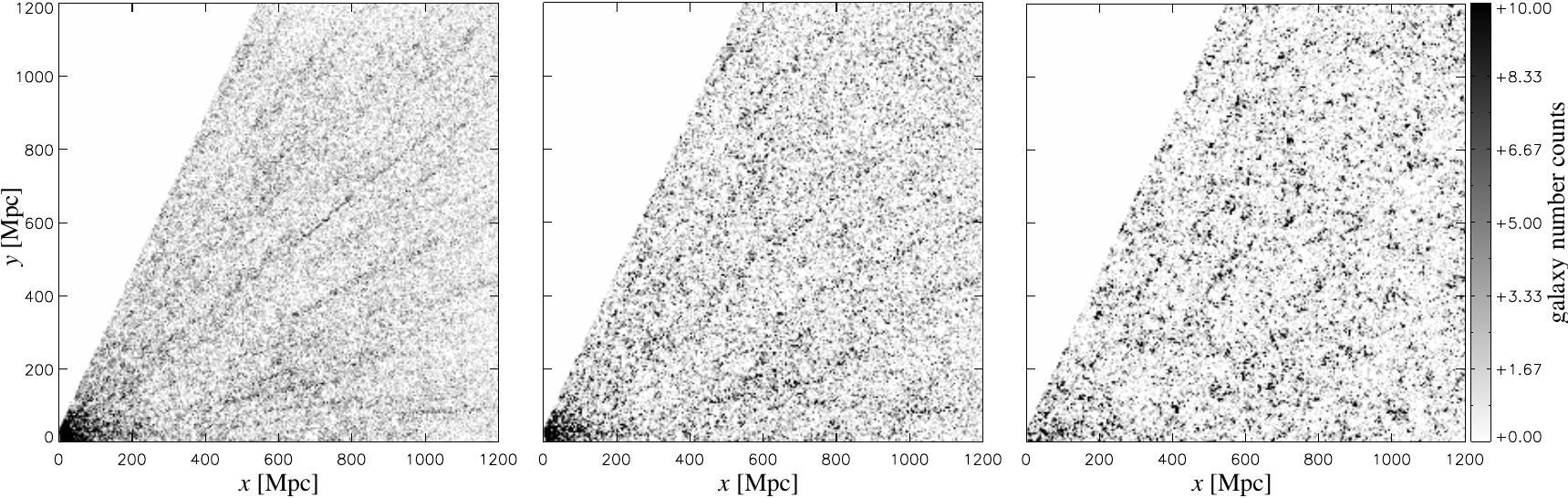}
\end{center}
\caption{Constrained realizations of the reconstructed density field. The data was simulated using an n-body simulation and the reconstruction assumes  the Poisson-lognormal prior with isotropic correlations. }
\label{fig:numdenssamples}       
\end{figure}

\begin{figure}[]
\begin{center}
\includegraphics[scale=0.8]{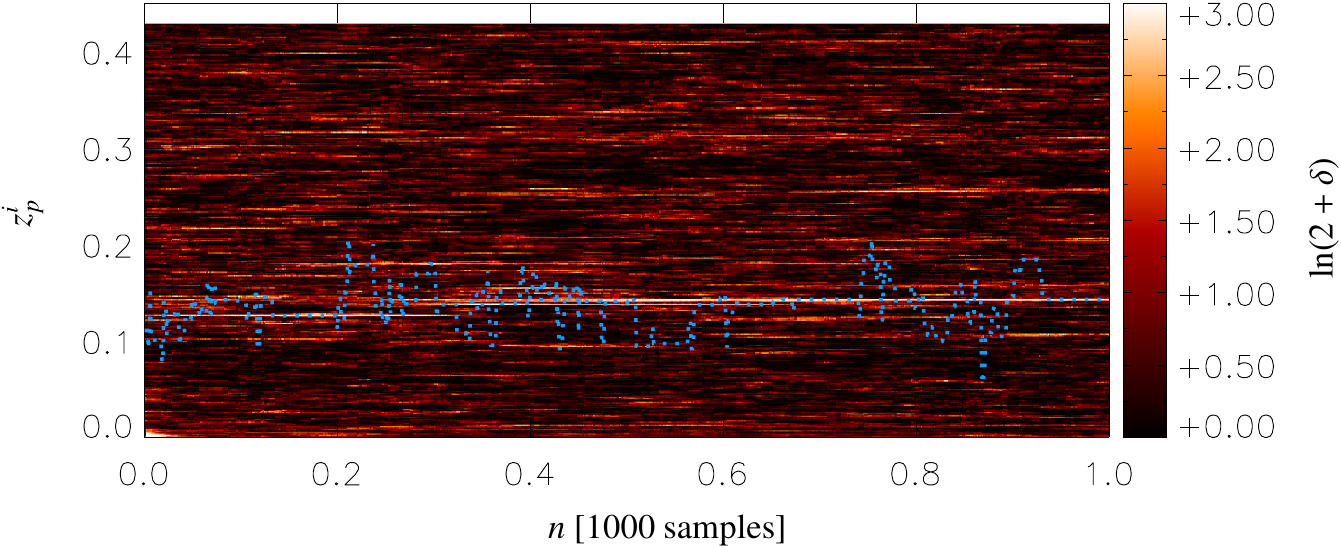}
\end{center}
\caption{Constrained realizations of the reconstructed density field. The data was simulated using an n-body simulation and the reconstruction assumes  the Poisson-lognormal prior with isotropic correlations. }
\label{fig:galaxywalk}       
\end{figure}

Figures~\ref{fig:numdenssamples} and \ref{fig:galaxywalk} illustrate  our approach. Even within a few steps the samples of the number density isotropize. As the sampler progresses, individual galaxies explore along their line of sight in a number density field which in turn fluctuates in response to the changing galaxy positions. 

Figures~\ref{fig:numdenssamples} and \ref{fig:galaxywalk} illustrate  our approach. The first figure shows that  even within a few steps the samples of the number density become isotropized. In the second figure we track the redshift of an example galaxy as the sampler explores the range of possible reconstructions. The galaxies explore along their line of sight in a number density field that, in turn, fluctuates in response to the changing galaxy positions. 

The results are  encouraging. In high density regions galaxy redshift uncertainties reduce by a factor of several. 
When a galaxy could reside in one of several concentrations lying along the line of sight the output pdf is multi-modal. 
Even so, the reconstructed redshift posteriors of the galaxies are generally far more informative than the inputs coming from photometric redshift estimators.   

\begin{figure}[]
\begin{center}
\includegraphics[scale=.55]{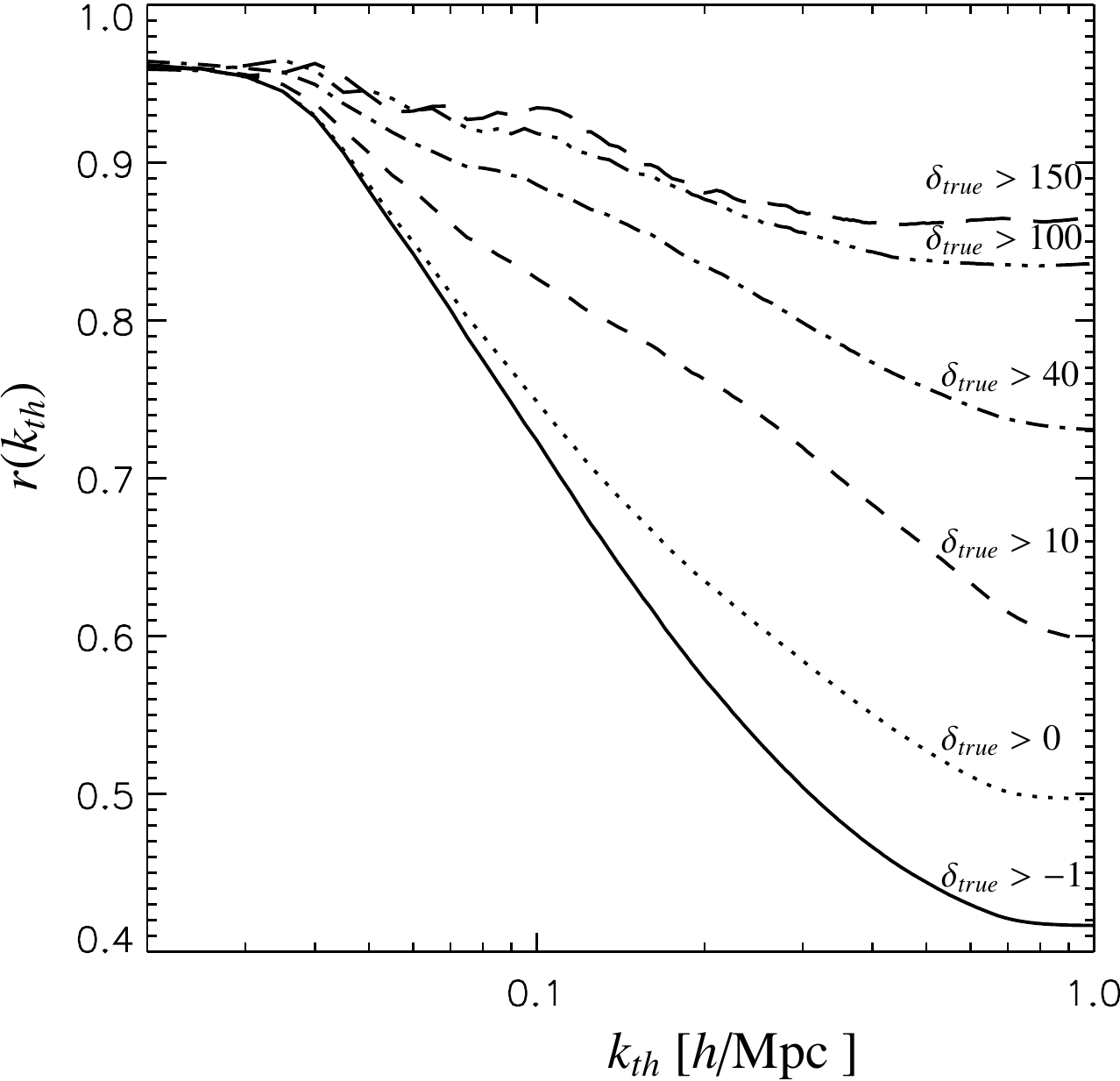}
\end{center}
\caption{The reconstructed density recovers the small scale features of the input density very well in high density regions. The figure shows the cross-correlation between the input field and the reconstructed density as a function of wave number. Different lines correspond to different thresholds of overdensity. }
\label{fig:r}       
\end{figure}

In order to summarize the result of the reconstruction we form the posterior mean estimator, the average of the number density field realizations that are explored by the sampler. We can compare this reconstruction to assess its capability to reproduce features of the input map. Figure~\ref{fig:r}  shows the k-space cross-correlations between the reconstructed and the input field. It is clear that the method is very successful in the high density parts of the sky.

\subsection{Discussion and conclusions}
\label{sec:discussion1}
The first main point of this talk is that we demonstrated the technical achievement of running a fully Bayesian analysis of a simulated data set with tens of millions of galaxies, and density fields represented on tens of millions of grid zones. The scale of this application corresponds to that of the current generation of available surveys, so it should be feasible to apply this approach to existing data.

The second key issue is to test whether our analysis is  sensitive to model misspecification, since the real data will not follow the correlated log-normal Poisson model. Our initial tests (of code correctness) used simulations that were consistent with the prior assumptions. These tests were passed. We do not show these tests here because the prior produces density fields that clearly not realistic, missing much of the filamentary structure which is characteristic of the cosmic web. 

The work we present in this talk  (and described in detail in Jasche \& Wandelt) uses simulated from an n-body simulation. Our results demonstrate that the reconstruction is successful in spite of using an approximate model.

The key feature underlying the reconstruction is clearly the ability to build in the prior assumption of isotropic correlations in the underlying cosmological number density field of the tracers. A secondary feature is the assumption of the shape of the correlations. What we show is that modeling those two aspects of the data results in acceptable reconstructions, that improve the redshift information for each galaxy significantly. It is also true that a better model including the morphological features of realistic gravitationally evolved number density would likely improve upon our results, since the differences between a correlated Poisson log-normal sample and a physical sample drawn from an n-body simulation are easily visible by eye. But it is clear that the reconstructions are not highly sensitive to the details of the assumed prior as long as two salient features of correlation and isotropy are included for the density field and we posit a simple statistical relationship of the tracers to the underlying density, in this case the inhomogeneous Poisson model.

Our approach is completely independent of and complementary to the means by which the photometric redshift is derived. The method is ready for tests on realistic data where the photoz pdfs will be  specified in terms of a different pdf for each galaxy.

As a consequence the method will be able to benefit from those tracers whose redshifts are better determined that others. In particular we  can merge the advantages of a large number of galaxies in photometric samples and the accuracy of spectroscopic samples! We will explore this idea further in follow-up studies.

This inference problem is of particular interest because it is an example where combining millions of noisy measurements with a physical prior, namely  the assumption of isotropic correlations produces a decisive gain in information.

In the second part of the talk we will see another application of the notion of statistical isotropy -- this time to the construction of an estimator for the expansion history of the Universe.

\section{Precision Cosmography with Cosmic Voids}
\label{sec:voids}\index{voids}
Understanding the physical properties of dark energy is a major goal of modern cosmology. There are essentially two distinct approaches to reaching this goal: cosmography and tracing structure formation.

\runinhead{Cosmography.}\index{cosmography}
The cosmography approach, which constrains dark energy properties using precision measurements of the expansion geometry of the Universe. Einstein's equation relates the geometrical properties of our Universe to its content. Since ``dark energy'' is just a placeholder for the terms in Einstein's equation that drive the observed accelerated  expansion of the Universe, precision cosmographical measurements can tell us about the time dependence of these terms and hence about the value, and rate of change of the equation of state parameter.

\runinhead{Tracing structure formation.}
The expansion of the universe has an impact on the rate at which primordial perturbations amplify. These perturbations then form structures through non-linear gravitational evolution, galaxy formation etc.. Observing the statistical properties (number, size etc) of these structures as a function of redshift constrains the growth of structure, and hence the expansion history, which is informative about the properties of dark energy. 

It is clear from this description that geometrical approaches are more direct. In addition, approaches relying on the statistical measures of the amount of structure in the universe inevitably require a detailed understanding of the processes that relate the formed structures to the underlying perturbation amplitude. These processes (e.g. galaxy formation) can be highly complex and deeply non-linear and are research areas in themselves. 

Geometrical approaches function by constructing standards out of observables (or combinations of observables) that can be modeled reliably such as standard candles (as in the case of type Ia supernovae), standard rulers (as in the case of Baryon Acoustic Oscillations (BAO)) or time standards (such as the (differences of) ages of galaxies). 

\subsection{The stacked voids Alcock-Paczynski test } \index{Alcock-Paczynski test}The Alcock-Paczynski (AP) test \cite{AP79} requires a different standard: ``standard, co-expanding spheres.'' One way to construct such standard spheres is through appealing to the statistical isotropy of the cosmological perturbations. In that case, correlations should depend only on the length, but not the direction of the vector connecting the two points being correlated. If the tracers that are being correlated did not move, any anisotropy in the correlation function could be interpreted as being due to the cosmological expansion at the redshift of the correlated objects. 

The key difficulty in constructing standard spheres are peculiar velocity effects. Any tracers that happen to lie in gravitationally bound structure will have velocities of the order of the depth of the gravitational potential well of the structure. For clusters or groups of galaxies the resulting finger of god effect in redshift space dominates the cosmic expansion signal by an order of magnitude. To construct an Alcock Paczyski test would therefore require a separate high precision measurement of the depth and shape of the potential well of any structures whose parts were used in the construction of the test.

So far the main work-around has been to only use very long range correlations of order 100 $h^{-1}$Mpc where peculiar velocity effects become sub-dominant compared to cosmic expansion effect and where the baryon sound speed at radiation drag leads to a peak in the correlation function. The downside of this limiting oneself to such large scales is that the statistical constraints will depend on the number of independent correlation volumes in survey volume, which limits the number of perturbation modes that can be used to arrive at the dark energy constraints and therefore leads one to consider extremely large surveys.

In this talk we propose a new way of constructing standard spheres: stacking cosmic voids. While  the AP test had been discussed for especially spherical individual voids
 \cite{Ryden95} stacking many voids guarantees spherical symmetry since isotropy prevents cosmic voids from pointing at us  (or away from us) preferentially. Finding voids in redshift shells, extracting them from the survey, co-centering them and stacking them, therefore gives rise to spherically symmetric underdensities. 

There are several advantages to using cosmic voids: 
\begin{itemize}
\item Voids are simple: peculiar velocities in and around voids are small compared to the cosmic expansion. We find that they give a 16\% systematic effect on our reconstructed Hubble diagram, with a very mild dependence on void size and redshift. 
\item Voids are small: A typical void size is 10 $h^{-1}$ Mpc - for a dense enough survey the number of voids per unit volume that can be detected is therefore of order 1000 times larger than the number of BAO correlation volumes. 
\item Voids remember: we find that voids have a well-ordered phase space - all they do is empty themselves out.
\end{itemize}
We use the term cosmic voids not to describe regions that are entirely empty, but regions that are underdense basins of repulsion in the cosmic density field. 

\begin{figure}[]
\begin{center}
\includegraphics[scale=.5]{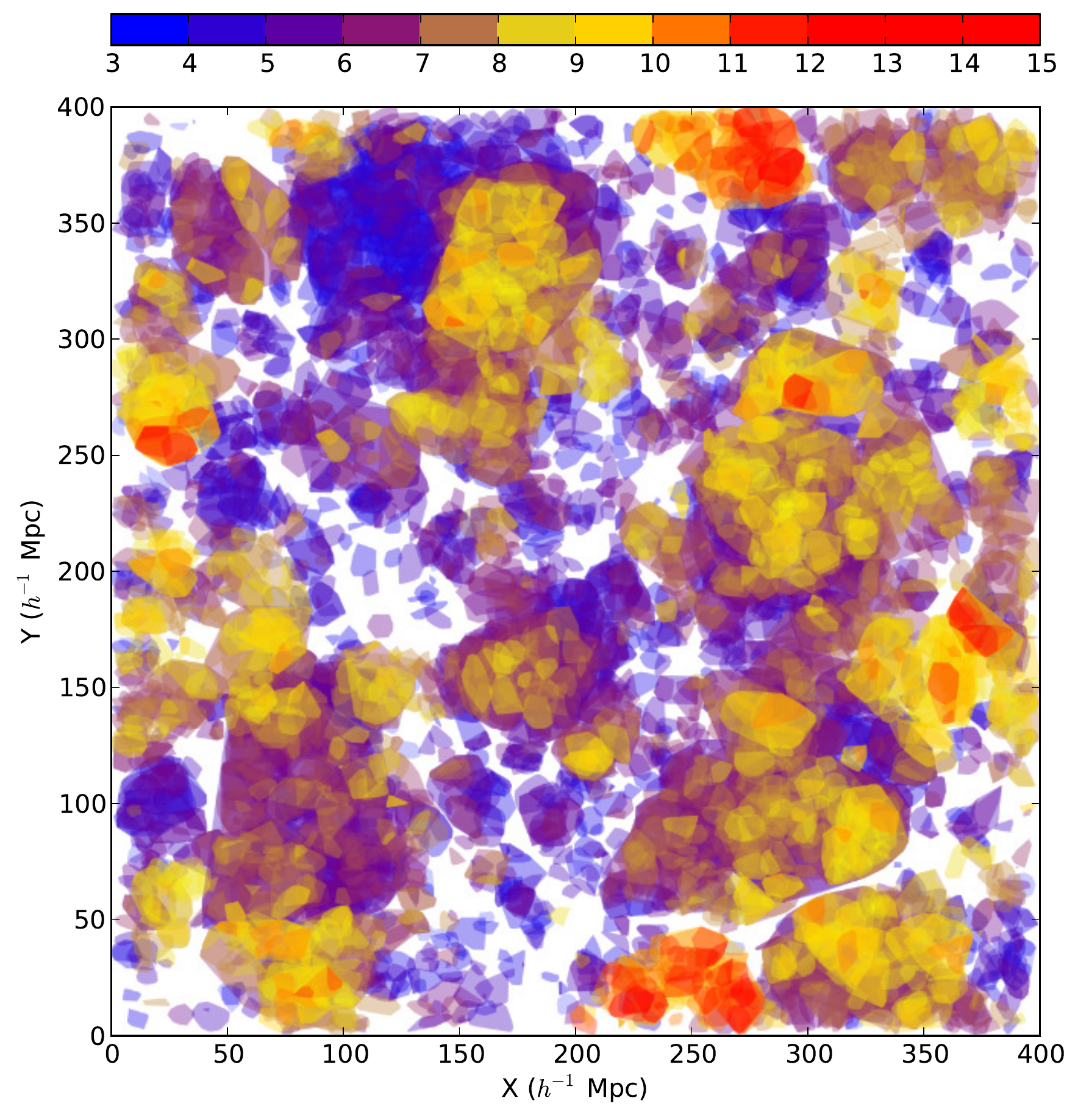}
\end{center}
\caption{The results of our void finder in a slice of an n-body simulation. The void finder constructs a hierarchical structure of voids. Each patch is a void, colored according to the level in the void hierarchy. When collecting voids in a size bin during the stacking procedure the algorithm traverses the tree in a depth first algorithm and marks and returns the first void it finds which satisfies the size criterion.  }
\label{fig:voidtree}       
\end{figure}

\begin{figure}[]
\begin{center}
\includegraphics[scale=.65]{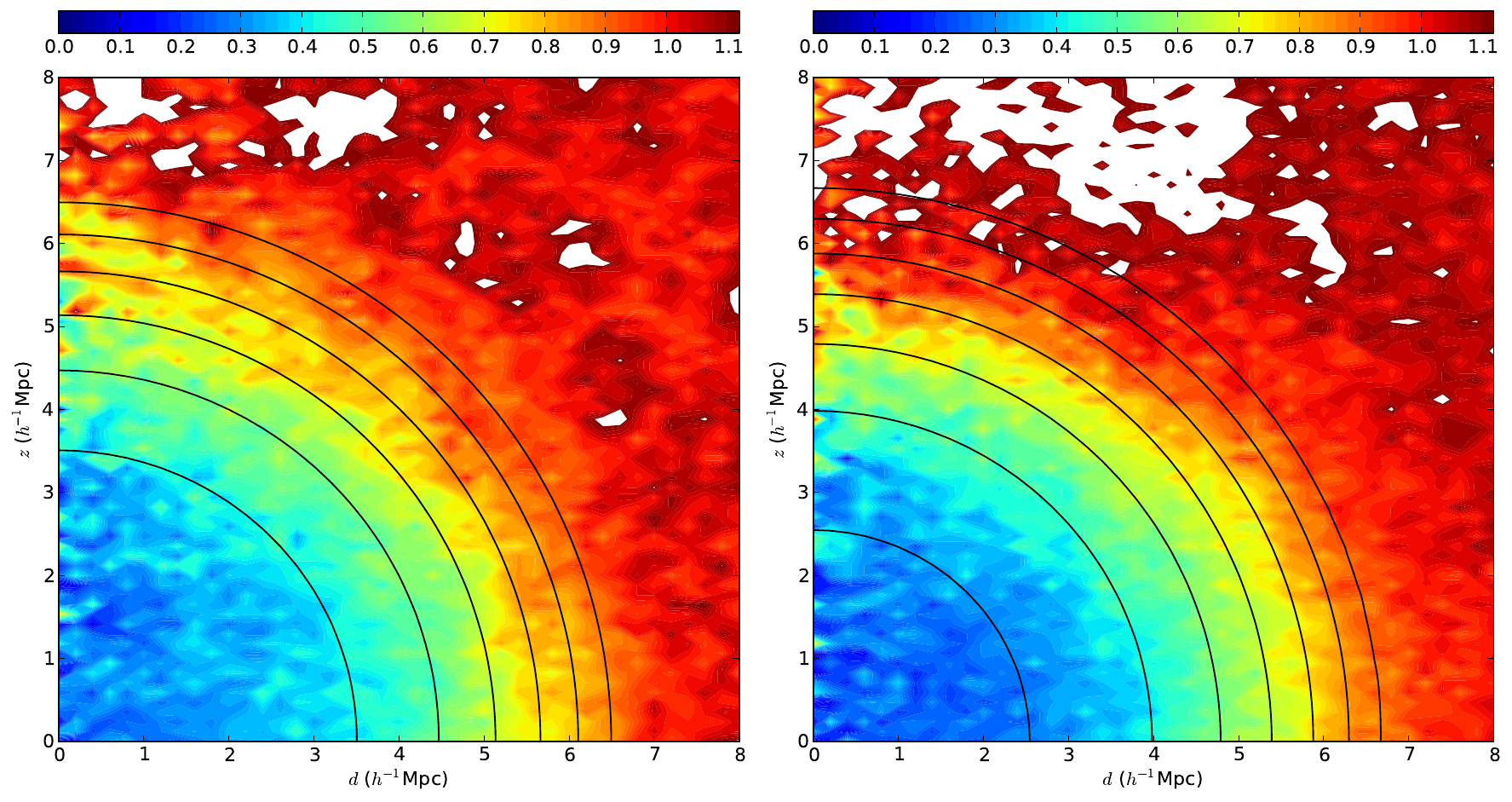}
\end{center}
\caption{A  void stack for 8 $h^{-1}$ Mpc voids. Left: the stack after fitting removing  the cosmic expansion effect, but without including peculiar velocities in the simulation. We find our profile agrees well with that found in \cite{WK93}. Right: The stack when  peculiar velocities are included. The same cosmic expansion has been removed as in the left panel. Careful inspection shows that peculiar velocities lead to a small net compression of the void stack along the line of sight.  }
\label{fig:voidstacks}       
\end{figure}

\begin{figure}[]
\begin{center}
\includegraphics[scale=.4]{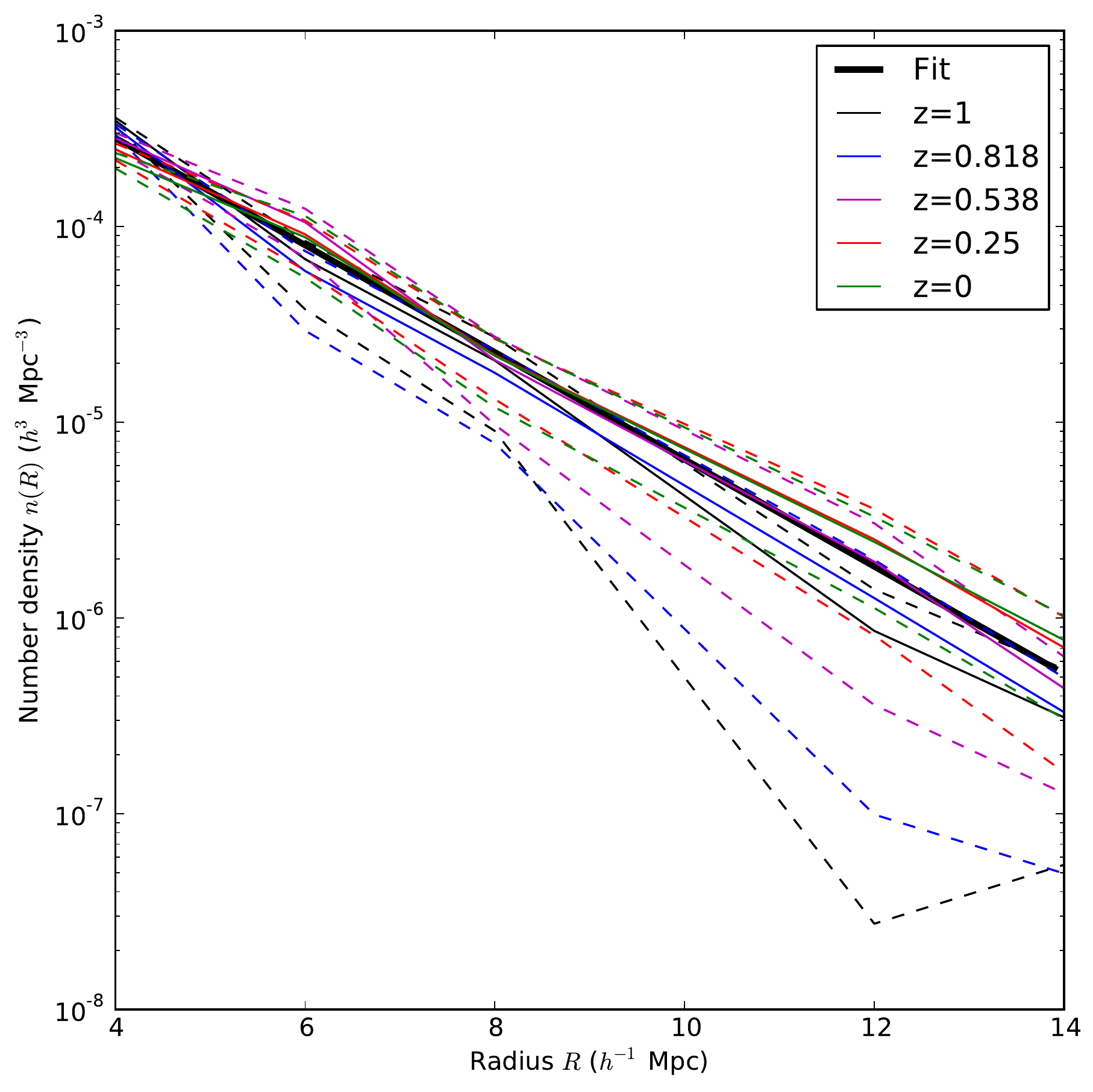}
\end{center}
\caption{Our simulation results for numbers densities of cosmic voids as a function of  redshift for voids of different sizes. These simulation results agree with the model described in \cite{SW04}.  }
\label{fig:voidnumber}       
\end{figure}

\begin{figure}[]
\begin{center}
\includegraphics[scale=0.65]{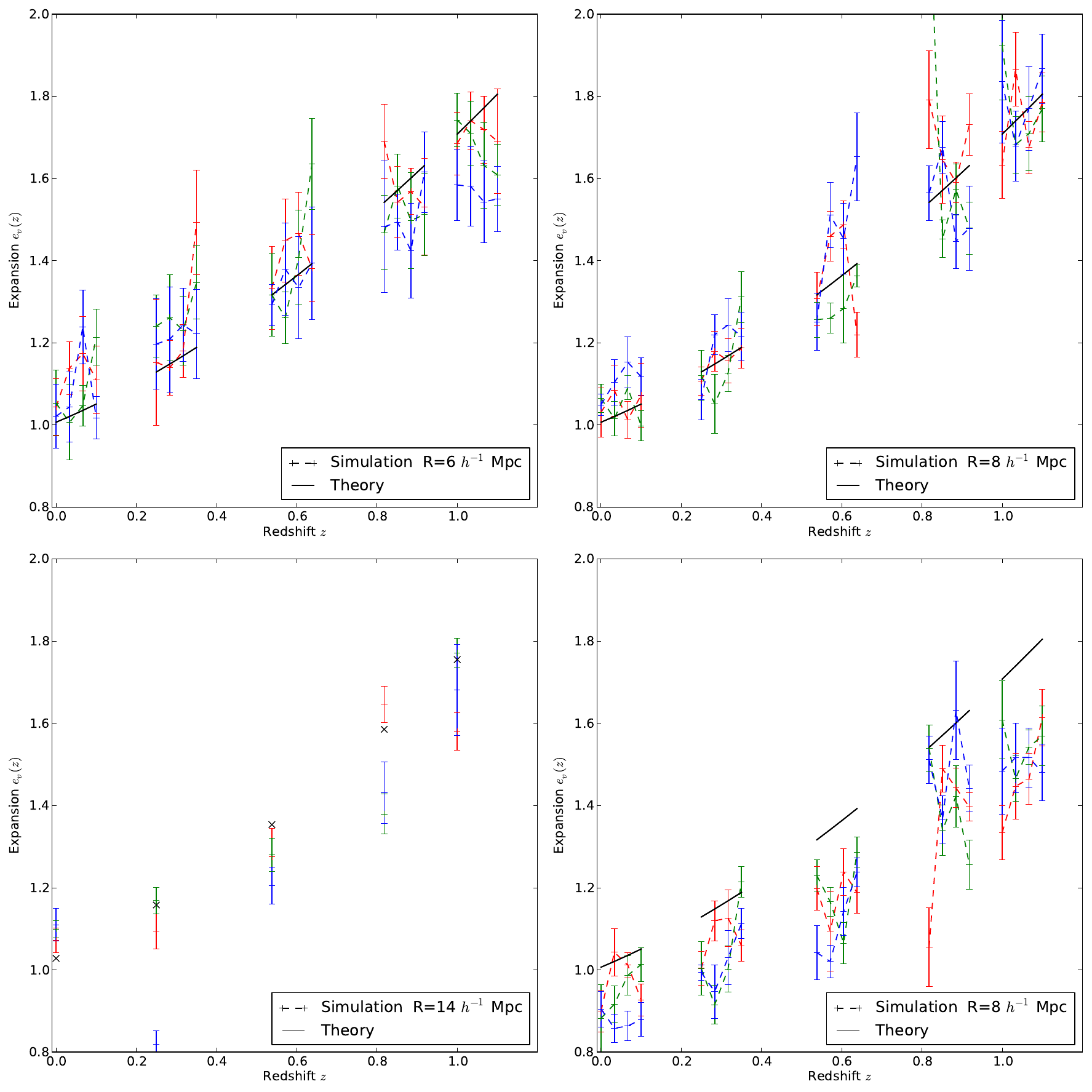}
\end{center}
\caption{The measured void stretch expansion history as a function of redshift for voids of 6, 8, and 14 $h^{-1}$Mpc (from left to right and top to bottom) for thhee simulations. The long-dashed line shows the result for the simulated cosmology. No peculiar velocities were included in the mock catalogs used for these plots. The lower right panel shows the result for 8 $h^{-1}$ Mpc voids for mocks with peculiar velocities and  without any correction for  peculiar velocity effect. The lack of redshift dependence of the resulting bias is clear. The same plot after debiasing is is shown in Figure~\ref{fig:debiasedstretch}.  }
\label{fig:stretch}       
\end{figure}

\begin{figure}[]
\begin{center}
\includegraphics[scale=0.5]{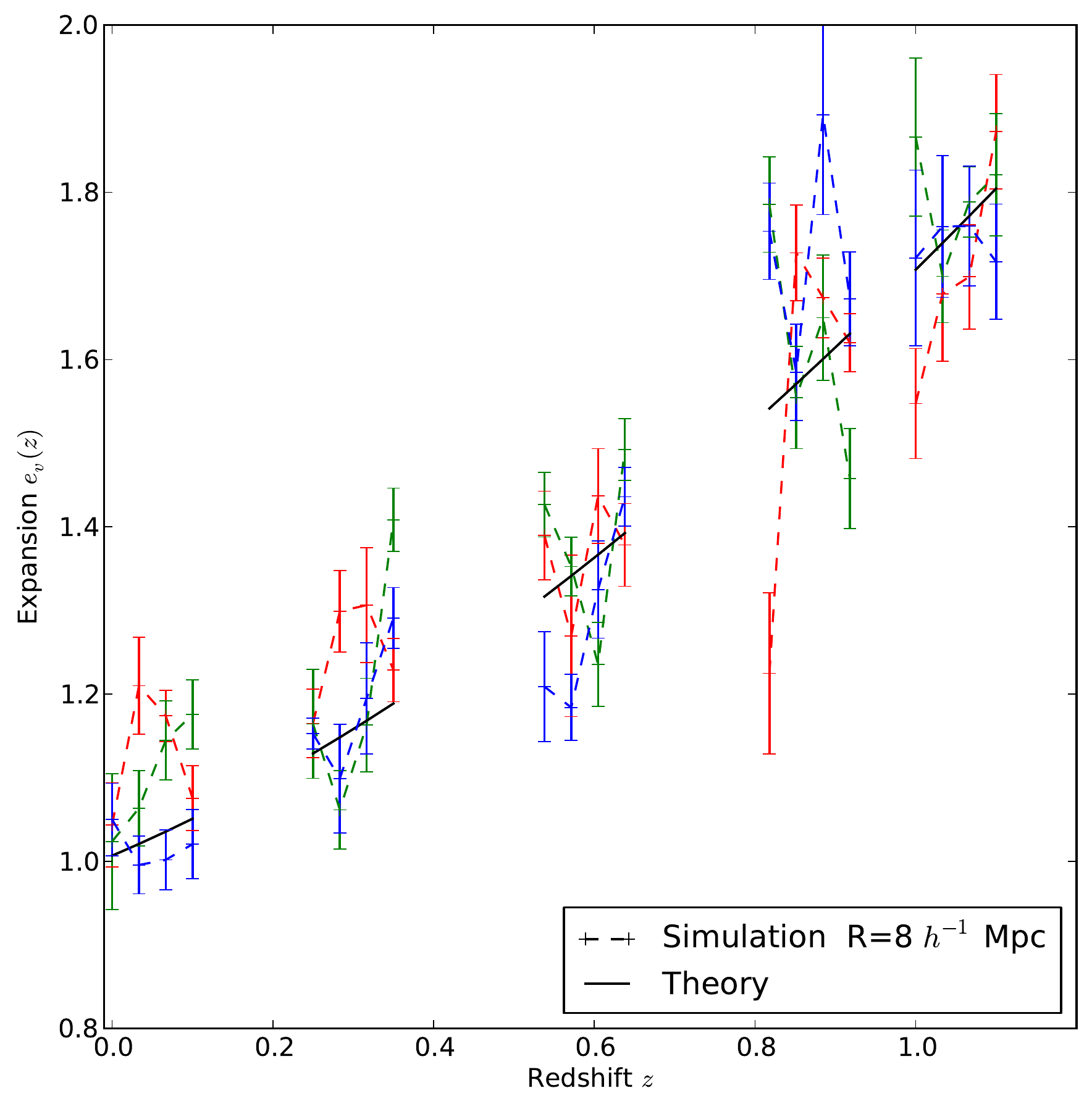}
\end{center}
\caption{Void stretch expansion history inferred from 8 $h^{-1}$ Mpc voids after the correction of a  peculiar velocity bias.  There is no evidence for  residual bias at the level of our simulations. }
\label{fig:debiasedstretch}       
\end{figure}

In order to demonstrate the promise of stacked voids for constructing a powerful AP test we solved the following problems:
\begin{enumerate}
\item create a suitable void definition: a modified  ZOBOV algorithm \cite{N08}  (see Figure~\ref{fig:voidtree});

\item define a method to add voids into stacks labeled by size and redshift, which both enhances signal to noise and sphericalizes them (see Figure~\ref{fig:voidstacks}); 
\item determine the number of voids that would be available to this method in an observed cosmological volume (see Figure~\ref{fig:voidnumber}); and
\item measure their stretch along the line of sight in order to obtain the expansion history of the universe (see Figures~\ref{fig:stretch} and Figure \ref{fig:debiasedstretch}).
\end{enumerate}
Details can be found in our main paper \cite{LW11}.

We   tested these methods in  a series of three pure dark matter $N$-body simulations with different realizations of the initial conditions but  the same cosmology. The volume of each simulation is given by a cube of side $L=500$\Mpch{}. Each simulation had $N=512^3$ particles. We  adopted a $\Lambda$CDM-WMAP7 cosmology with the 
following parameters: $\Omega_\mathrm{b} h^2 = 0.02258$, $\Omega_\mathrm{c} h^2=0.1108$, $H = 71$~\kmsMpc, $w=-1$, $n_\mathrm{S}=1$, $A_\mathrm{S}=2.34\times 10^{-9}$. This corresponds to 
$\Omega_\mathrm{b} = 0.045$, $\Omega_\mathrm{M}=0.264$, $\sigma_8=0.84$. Each particle had a mass $m_\mathrm{p}=2.05\,10^{11}$\hMsun.  The transfer function for density fluctuations for this cosmology was computed using {\sc CAMB} \cite{CAMB}. The initial conditions were generated using {\sc ICgen},\footnote{Available from http://www.iap.fr/users/lavaux/.} a code which uses the transfer function to generate a density field from the primordial power spectrum.

\subsection{Discussion and Conclusion}
Based on these results we performed a Fisher matrix forecast of the statistical constraints on dark energy equation of state parameter $w_{a}$ and its rate of change $w_{p}$  that we would expect from Euclid. We quantify the answer in terms of the figure of merit defined by the Dark Energy Task Force \cite{detf}, ie the relative reduction in the area of the uncertainty ellipse for these two quantities. The result is exciting--we find that the stacked void Alcock-Paczynski test has the potential significantly to enhance the power of the proposed (and now selected) Euclid space craft  to constrain dark energy phenomenology. 

On the face of it cosmic voids have the potential to provide a far more powerful constraint on dark energy than measurements of the Baryonic Acoustic Oscillation scale, by up to an order of magnitude. This large increase of information is easily understood in comparing the number of modes probed by voids compared to BAOs, which scales roughly as the third power of the ratio of the BAO scale to the scale of the smallest usable voids $\sim 1000$. The area of parameter constraints scales as the square root of the number of modes $\sim 30$. When projected into the $w_{a},w_{p}$ plane using the Fisher matrix formalism for the EUCLID wide survey, we find the improvement over BAO on those parameters by a factor of $\sim 10$. 

We expect our stacked void shape measurements  to be robust to galaxy bias as it is purely geometrical and relies on the topology of the density field \cite{S98}. In fact, it is possible that biased tracers of the density enhance the contrast of voids and therefore enhance the void detection rate. These expectations remains to verified on more realistic mock catalogs and real data. 

Based on our Fisher matrix forecasts, the stacked voids technique promises a remarkable increase to the figure of merit from EUCLID  when compared to the combined results from all other probes using EUCLID data (BAO, weak lensing, type Ia supernovae, cluster counts). The Alcock-Paczinsky  test using stacked voids is therefore potentially a significant addition to the portfolio of major dark energy probes which merits further detailed  studies focused on additional real-world systematics and optimal survey design.

\begin{acknowledgement}
 The authors wish to thank Eric Feigelson and Jogesh Babu for the invitation to speak at this highly enjoyable and fruitful meeting. We thank Laird Thompson, Joseph Silk, Mark Neyrinck, Thierry Sousbie, Miguel Arag\'on-Calvo, and St\'ephane Colombi  for useful discussions and we owe a special debt of gratitude to the respondent, Christopher Genovese for his insightful and valuable commentary.   

The authors acknowledge financial support from NSF Grants AST 07-08849, AST 09-08693 and from BDW's Chaire d'Excellence granted by the Agence Nationale de Recherche. GL acknowledges support from CITA National Fellowship and financial support from the Government of Canada Post-Doctoral Research Fellowship. This research was supported by the National Science Foundation through TeraGrid resources provided by NCSA under grant number TG-AST100029.

We thank Francisco S. Kitaura, Torsten A. En\ss lin and Simon D. White for useful discussions; Cristiano Porciani for the simulated density field and Nina Roth for help with handling the simulation data; Rainer Moll and Bj\"{o}rn M. Sch\"{a}fer for useful discussions and support with many valuable numerical gadgets. This work has been supported by the Deutsche Forschungsgemeinschaft within the Priority Programme 1177 under the project PO 1454/1-1.

Research at Perimeter Institute is supported by the Government of Canada through Industry Canada and by the Province of Ontario through the Ministry of Research and Innovation.

\end{acknowledgement}
%
%
%

%
%
%
\biblstarthook{}

\end{document}